\documentstyle[prd,aps,preprint,eqsecnum,psfig,floats]{revtex}
\def\bege{\begin{equation}}
\def\ende{\end{equation}}

\def\bar#1{\overline{ #1 }}

\def\beq{\begin{equation}}
\def\eeq{\end{equation}}
\def\bea{\begin{eqnarray}}
\def\eea{\end{eqnarray}}

\begin{document}
\tighten

\title{Poincar{\'e} Invariant Algebra From Instant to Light-Front 
Quantization}

\author{Chueng-Ryong Ji and Chad Mitchell}

\address{Department of Physics, North Carolina State University,
Raleigh, NC 27695-8202, USA}

%\date{DRAFT: \today}

\maketitle

\begin{abstract}

We present the Poincar{\'e} algebra interpolating between
instant and light-front time quantizations. The angular momentum 
operators satisfying SU(2) algebra are constructed in an arbitrary
interpolation angle and shown to be identical to the ordinary
angular momentum and Leutwyler-Stern angular momentum in the instant 
and light-front quantization limits, respectively.
The exchange of the dynamical role between the transverse angular mometum 
and the boost operators is manifest in our newly constructed algebra.
 
\end{abstract}
\pacs{ }

\section{Introduction}

When hadronic systems are described in terms of quarks and gluons, 
it is part of nature that the characteristic momenta
are of the same order or even very much larger than the masses of the
particles involved. For example, relativistic effects are crucial to 
describe the low-lying hadrons made of $u,d$ and $s$ quarks and 
anti-quarks\cite{Isgur2}. It has also been realized that a 
parametrization of nuclear reactions in terms of non-relativistic wave 
functions must fail. Thus, a relativistic treatment is 
one of the essential ingredients that should be incorporated in 
developing a successful strong interaction theory.

For the relativistic Hamiltonian approach, several forms of 
dynamics have been suggested\cite{Dirac,6}. Although the point form dynamics 
has also been explored recently\cite{Austria}, the most popular choices
were thus far the equal-$t$ (instant form) and equal-$\tau=t+z/c$ 
(light-front form) quantizations. A crucial difference between the 
instant form and the light-front form may be attributed to their 
energy-momentum dispersion relations. When a particle has the mass $m$ 
and the four-momentum $k = (k^{0},k^{1},k^{2},k^{3})$, the relativistic 
energy-momentum dispersion relation of the particle at equal-$t$ is given by
\begin{equation} \label{droft}
k^{0} = \sqrt{\vec{k}^{2} + m^{2}},
\end{equation}
where the energy $k^{0}$ is conjugate to $t$ and the three-momentum
vector $\vec{k}$ is given by $\vec{k} = (k^{1},k^{2},k^{3})$.
However, the corresponding energy-momentum dispersion relation
at equal-$\tau$ is given by
\begin{equation} \label{droftau}
k^{-} = \frac{\vec{k}_{\bot}^{2} + m^{2}}{k^{+}},
\end{equation}
where the light-front energy conjugate to $\tau$ is given by
$k^{-} = k^{0} - k^{3}$ and the light-front momenta $k^{+} =
k^{0} + k^{3}$ and $\vec{k}_{\bot} = (k^{1},k^{2})$ are orthogonal
to $k^{-}$ and form the light-front three-momentum $\underline{k} =
(k^{+},\vec{k}_{\bot})$.
While the instant form (Eq.(\ref{droft})) exhibits an irrational 
energy-momentum relation, the light-front form (Eq.(\ref{droftau})) 
yields a rational relation and thus
the signs of $k^{+}$ and $k^{-}$ are correlated,
{\it e.g.} the momentum $k^{+}$ is always positive 
when the system evolve to the future direction 
({\it i.e.} positive $\tau$)) so that
the light-front energy $k^{-}$ is positive. 
In the instant form, however, no sign 
correlations for $k^{0}$ and $\vec{k}$ exist. 
Such a dramatic difference in the
energy-momentum dispersion relation makes the light-front quantization 
quite distinct from other forms of the Hamiltonian dynamics.

The light-front quantization \cite{Dirac,Steinhardt} has already been 
applied successfully in the context of current algebra \cite{lcca}
and the parton model \cite{lcparton} in the past.
With the recent advances in the Hamiltonian renormalization 
program\cite{brodsky,wilson}, Light-Front Dynamics (LFD) appears to be 
even more promising for the relativistic treatment of hadrons.
In the work of Brodsky, Hiller and McCartor \cite{Hil},
it is demonstrated how to solve the problem of renormalizing light-front
Hamiltonian theories while maintaining Lorentz symmetry and other
symmetries. The genesis of the work presented in \cite{Hil} may be found in
\cite{RM} and additional examples including the use of LFD methods to
solve the bound-state problems in field theory can be found in the 
recent review\cite{BPP}. These results are indicative of the great 
potential of LFD 
for a fundamental description of non-perturbative effects in strong 
interactions. This approach may also provide a bridge between the two
fundamentally different pictures of hadronic matter, i.e. the
constituent quark model (CQM) (or the quark parton model) closely
related to experimental observations and the quantum chromodynamics
(QCD) based on a covariant non-abelian quantum field theory.
Again, the key to possible connection between the two pictures is the
rational energy-momentum dispersion relation given by Eq.(\ref{droftau})
that leads to a relatively simple vacuum structure. There is no 
spontaneous creation of massive fermions in the LF quantized vacuum. 
Thus, one can immediately obtain a constituent-type 
picture~\footnote{To provide further insight concerning this issue,
we have recently introduced an IR longitudinal cutoff and generated a 
light-front counterterm which sets a scale for a dynamical mass gap for 
quarks and gluons as well as a string tension in the light-front
QCD Hamiltonian\cite{GJC}.}, in which all 
partons in a hadronic state are connected directly to the hadron instead 
of being simply disconnected excitations (or vacuum fluctuations) in a 
complicated medium. A possible realization of chiral symmetry breaking in 
the LF vacuum has also been discussed in the literature~\cite{Wilson}.

Furthermore, one of the most popular formulations for the analysis of
exclusive processes involving hadrons exists in the framework of
light-front (LF) quantization~\cite{BPP}.
In particular, the Drell-Yan-West ($q^+=q^0+q^3=0$) frame
has been extensively used in the calculation
of various electroweak form factors and decay
processes~\cite{Ja2,CJ1,Kaon}. 
In this frame\cite{DYW}, one can derive a first-principle
formulation for the exclusive amplitudes by 
choosing judiciously the component of the light-front current.
As an example, only the parton-number-conserving (valence) Fock state 
contribution is needed in $q^+=0$ frame when the ``good" component of the 
current, $J^+$ or ${\bf J}_{\perp}=(J_x,J_y)$, is used for the spacelike
electromagnetic form factor calculation of pseudoscalar mesons.
One doesn't need to suffer from complicated vacuum
fluctuations in the equal-$\tau$ formulation 
once again due to the rational dispersion relation. 
The zero-mode contribution may also be
avoided in Drell-Yan-West frame by using the plus component of current
\cite{Ji1}.
However, caution is needed in applying the established
Drell-Yan-West formalism to other frames because the current components
do mix under the transformation of the reference-frame\cite{chad}.

In LFD a Fock-space expansion of bound states is made. The wave
function $\psi_n(x_i, k^\perp_i, \lambda_i)$ describes the
component with $n$ constituents, with longitudinal momentum fraction
$x_i$, perpendicular momentum $k^\perp_i$ and helicity $\lambda_i$,
$i=1, \dots, n$. It is the aim of LFD to determine those wave functions
and use them in conjunction with hard scattering amplitudes to describe
the properties of hadrons and their response to electroweak probes.
Important steps were taken towards a realization of
this goal\cite{Hil}.
However, at present
there are no realistic results available for wave functions of hadrons
based on QCD alone. In order to calculate the response of hadrons to external
probes, one might resort to the use of model wave functions.
This way to estimate matrix elements has been presented in many literatures
\cite{teren,Dziem,Ji,choi,Jaus1,Jaus,Chung,choi1,Huang,schlumpf,card}.
Especially, the variational principle enabled the solution of a 
QCD-motivated effective Hamiltonian, and the constructed LF quark-model 
provided a good description of the available experimental data spanning 
various meson properties \cite{ChoiJi}.
The same reasons that make LFD so attractive to solve
bound-state problems in field theory make it also useful for a
relativistic description of nuclear systems.  LF methods have the
advantage that they are formally similar to time-ordered many-body
theories, yet provide relativistically invariant observables.

On the other hand, the Poincar{\'e} algebra in the ordinary equal-$t$ 
quantization is drastically changed in the light-front equal-$\tau$ 
quantization.
Although the maximum number (seven) of the ten Poincare
generators are kinematic (i.e. interaction independent) and they leave
the state at $\tau = t + z/c =0$ unchanged~\cite{5},
rotation becomes a dynamical problem in the light-front quantization. 
Because the quantization surface $\tau$ = 0 is not invariant
under the transverse rotation whose direction is perpendicular to the
direction of the quantization axis $z$ at equal $\tau$\cite{surya}, 
the transverse angular momentum operator involves the interaction
that changes the particle number. 
Leutwyler and Stern showed that the angular momentum operators can be 
redefined to satisfy the SU(2) spin algebra and the commutation relation 
between mass operator and spin operators~\cite{6};
\begin{equation}
[{\cal{J}}_i,{\cal{J}}_j] = i {\epsilon_{ijk}} {\cal{J}}_k,
\end{equation}
\begin{equation}
[M,{\cal{\vec J}}] =0.
\end{equation}
However, in LFD, there are two dynamic equations to solve:
\begin{equation}
{\cal J}^2 |H; p^+, {\vec p_{\perp}}^2> =
{S_H}({S_H}+1) |H; p^+, {\vec p_{\perp}}^2>
\end{equation}
and
\begin{equation}
M^2 |H; p^+, {\vec p_{\perp}}^2> =
{m_H}^2 |H; p^+, {\vec p_{\perp}}^2>,
\end{equation}
where the total angular momentum(or spin) and the
mass eigenvalues of the hadron($H$) are given by
$S_H$ and $m_H$.
Thus, it is not a trivial matter to specify
the total angular momentum of a specific hadron state.

As a step towards understanding the conversion of the dynamical problem
from boost to rotation, in this work we construct the Poincar{\'e} 
algebra interpolating between instant and light-front time quantizations.
We use an orthogonal coordinate system which interpolates smoothly 
between the equal-time and the light-front quantization hypersurface. 
Thus, our interpolating coordinate system has a nice feature of tracing 
the fate of the Poincare algebra at equal time as the hypersurface 
approaches to the light-front limit.
The same method of interpolating hypersurfaces has been used by Hornbostel
\footnote{Application to the axial anomaly in the Schwinger model has also
been presented\cite{JiRey}.}.
In an arbitrary interpolation angle, we find the transformation that allows
not only the simultaneous assignments of mass and angular momentum but
also SU(2) algebra among the angular momentum operators. Approaching 
the light-front limit, we verify that the LFD has one more kinematic 
operator than the dynamics with any other interpolation angle.
Also, we find that the roles of angular momentum and boost are smoothly
exchanged as the interpolation angle moves from $t$ to $\tau$.
We also obtain a general definition of ${\cal J}_\perp$ and ${\cal J}_3$
at an arbitrary interpolation angle and show that it is consistent
with the result
obtained by Leutwyler and Stern in the light-front limit.

In the next section, Section II, we present the Poincar{\'e} algebra
interpolating between equal-$t$ and equal-$\tau$. In Section III, we
construct the angular momenta that satisfy the SU(2) spin algebra in any
interpolation angle and present the two dynamic equations to be solved 
simultaneously in an arbitrary interpolation angle. Discussion of results
and conclusions follow in Section IV. In Appendix A, we summarize
the forty-five commutation relations for the Poincar{\'e} generators
with an arbitrary interpolation angle. In Appendix B, we provide explicit
representations of the helicity operator and the spin 1 and spin 1/2 polarization
vectors with an arbitrary interpolation angle.

\section{Interpolation Angle Dependent Poincar{\'e} Algebra}

We begin by introducing an interpolating parameter $\delta$.  Previous authors
have used the parameter $\frac{\pi}{2}\le\theta\le\pi$ such that
\begin{eqnarray}
\left(\begin{array}{cc} x^{+} \\ x^{-}
\end{array} \right) = 
\left(\begin{array}{cc} \sin{\frac{\theta}{2}} &
\cos{\frac{\theta}{2}} \\ \cos{\frac{\theta}{2}} & -\sin{\frac{\theta}{2}} 
\end{array} \right)\left(\begin{array}{cc} x^{0} \\ x^{3}
\end{array} \right).
\end{eqnarray}
Here $x^+$ plays the role of "time" and $x^-$ is the longitudinal coordinate
as defined on an arbitrary interpolation front.
In this work we define $\delta = \frac{\pi}{2} - \frac{\theta}{2}$ so that
\begin{eqnarray}
\left(\begin{array}{cc} x^{+} \\ x^{-}
\end{array} \right) = 
\left(\begin{array}{cc} \cos{\delta} &
\sin{\delta} \\ \sin{\delta} & -\cos{\delta} 
\end{array} \right)\left(\begin{array}{cc} x^{0} \\ x^{3}
\end{array} \right).
\end{eqnarray}
This parameter is easily visualized and ranges from $\delta=0$ on the equal-time
instant $x^0=0$ to $\delta=\frac{\pi}{4}$ on the light-front 
${x^+}=\frac{1}{\sqrt{2}}({x^0}+{x^3})=0$.
In this new basis the metric becomes
\begin{equation}
[g_{\mu\nu}] = 
\left(\begin{array}{cccc} C & 0 & 0 & S \\ 0 & -1 & 0 & 0 \\ 0 & 0 & -1 & 0 \\ S
& 0 & 0 & -C \end{array} \right),
\end{equation}
where $C = \cos{2\delta}$, $S = \sin{2\delta}$ and $g_{++}=C$.
Similarly, we transform the Poincar{\'e} matrix to this new basis, so that
\begin{equation}
[M^{\mu\nu}] =
\left(\begin{array}{cccc} 0 & K^1 & K^2 & K^3 \\ -K^1 & 0 & J^3 & -J^2 \\ -K^2
& -J^3 & 0 & J^1 \\ -K^3 & J^2 & -J^1 & 0 \end{array} \right)
\rightarrow
[M^{\mu\nu}] =
\left(\begin{array}{cccc} 0 & E^1 & E^2 & -K^3 \\ -E^1 & 0 & J^3 & -F^1 \\ -E^2
& -J^3 & 0 & -F^2 \\ K^3 & F^1 & F^2 & 0 \end{array} \right),
\end{equation}
where we introduce the operators
\begin{eqnarray}
E^1 = J^2\sin{\delta} + K^1\cos{\delta} \nonumber \\
E^2 = K^2\cos{\delta} - J^1\sin{\delta} \nonumber \\
F^1 = K^1\sin{\delta} - J^2\cos{\delta} \nonumber \\
F^2 = J^1\cos{\delta} + K^2\sin{\delta} 
\end{eqnarray}
on an arbitrary interpolation front.

The ten generators of the Poincar{\'e} algebra are 
${P_{+},P_{-},P_1,P_2,E_1,E_2,F_1,F_2,K_3,J_3}$, where
each Poincar{\'e} generator is defined on the interpolation front as follows.
The Hamiltonian becomes $P_{+} = CP^{+} + SP^{-} = AP^0 + BP^3$, where
\begin{eqnarray}
A = C\cos{\delta}+S\sin{\delta}, \nonumber \\
B = C\sin{\delta}-S\cos{\delta}. 
\end{eqnarray}
Similarly the momentum vector is $(P_{-},\bf {P_{\perp}})$ where $P_{-}=S{P^+}-C{P^-}
=-B{P^0}+A{P^3}$.  The transverse rotation operators can be read from
$[M^{\mu\nu}]$ to be $F_1$ and $F_2$.  As in both the equal-time and light-front
cases, transverse rotations are chosen to commute with the Hamiltonian, $[F_{i},P_{+}] = 0$.
Finally, the transverse boost operators for an arbitrary interpolation front are
$E_1$ and $E_2$.  Again as in both the equal-time and light-front cases,
transverse boosts are chosen to commute with the longitudinal momentum, $[E_i,P_{-}] = 0$.
Note that the longitudinal angular momentum and longitudinal boost operators are
essentially unaffected by the transformation to an arbitrary interpolation
front.

Other commutation relations among the ten generators may be obtained from the 
usual rules $[M^{\rho\sigma},P^{\mu}] = -i(g^{\mu\rho}-g^{\mu\sigma}P^{\rho})$
and $[M^{\alpha\beta},M^{\rho\sigma}] =
-i(g^{\beta\sigma}M^{\alpha\rho}-g^{\beta\rho}M^{\alpha\sigma}+
g^{\alpha\rho}M^{\beta\sigma}-g^{\alpha\sigma}M^{\beta\rho})$.  A comprehensive
list of the 45 commutation relations among the contravariant components of the
Poincar{\'e} generators is presented in the Appendix.  This algebra is consistent with
the equal-time algebra for $\delta=0$; it is also consistent with the
light-front algebra for $\delta=\frac{\pi}{4}$.

Next we investigate the algebraic structure of the Poincar{\'e} group on an
arbitrary interpolation front.  The stability group of the initial surface
$x^+=0$ is the set of operators which generate Poincar{\'e} transformations that
leave this surface invariant.  Following the literature, we describe such
operators as kinematical.  In physical terms, kinematical operators are those
operators that do not change the direction of the time ($x^+$) axis.  To
clarify the distinction between kinematic and dynamic operators, we define
an alternate set of Poincar{\'e} genarators by transforming the Poincar{\'e} 
matrix:  $M_{\alpha\beta} =  {g_{\alpha\mu}}{g_{\beta\nu}}M^{\mu\nu}$, so that:
\begin{equation}
[M_{\alpha\beta}] =
\left(\begin{array}{cccc} 0 & {\cal D}_1 & {\cal D}_2 & K^3 \\
 -{\cal D}_1 & 0 & J^3 & -{\cal K}_1 \\ -{\cal D}_2 & -J^3 & 0 & -{\cal K}_2 \\
  -K^3 & {\cal K}_1 & {\cal K}_2 & 0 \end{array} \right),
\end{equation}
where the new generators are defined as follows:
\begin{eqnarray}
{\cal K}_1 &=& CF_1 - SE_1 \nonumber \\
{\cal K}_2 &=& CF_2 - SE_2 \nonumber \\
{\cal D}_1 &=& -SF_1 - CE_1 \nonumber \\
{\cal D}_2 &=& -SF_2 - CE_2. 
\end{eqnarray}

It can be seen that $[{\cal K}_i,P^{+}] = 0$, and therefore each 
transformation $e^{-i\omega{\cal K}_i}$ leaves the $P^{+}$ operator
invariant.  Thus the $p^{+}$ eigenvalue for a given momentum state is
invariant under $e^{-i\omega{\cal K}_i}$.  It follows that the +
component of any four-vector is invariant under $e^{-i\omega{\cal K}_i}$,
and therefore $e^{-i\omega{\cal K}_i}|x^{+}>\sim{|x^{+}>}$.  As the instant
$x^{+} = 0$ is unaltered, ${\cal K}_1$ and ${\cal K}_2$ are kinematic.

The sets of kinematic and dynamic generators are presented in Table 1.

\begin{table*}
\begin{center}
\caption{Kinematic and dynamic generators on an arbitrary interpolation
front.}
\begin{tabular}{|c|c|c|}
&Kinematic & Dynamic\\
\hline
$\delta=0$ & ${\cal K}_1=-J_2$, ${\cal K}_2=J_1$, $J_3$, $P^1,P^2,P^3$ & 
${\cal D}_1=-K_1$, ${\cal D}_2=-K_2$, $K_3$, $P^0$ \\
\hline
$0\le\delta{<}\frac{\pi}{4}$ & ${\cal K}_1$, ${\cal K}_2$, $J_3$, $P^1,P^2,P_{-}$ & 
${\cal D}_1$, ${\cal D}_2$, $K_3$, $P_{+}$ \\
\hline
$\delta=\frac{\pi}{4}$ & ${\cal K}_1=-E_1$, ${\cal K}_2=-E_2$, $J_3$, $K_3$, $P^1$,
$P^2$,$P^{+}$ & 
${\cal D}_1=-F_1$, ${\cal D}_2=-F_2$, $P^{-}$ \\
\end{tabular}
\end{center}
\end{table*}

There are two key features to be noted in this algebra.  The first is
the appearance of the longitudinal boost operator in the stability
group on the light-front.  The number of kinematic generators remains 
unchanged until we reach the light-front quantization, where the operator
$K^3$ becomes kinematic.  To understand this, note that $[P^+,K^3] = iP_{-} = 
i(SP^+ - CP^-) \rightarrow iP^+$ as $\delta\rightarrow 
\frac{\pi}{4}$.  Similarly we have $[x^+,K^3] = ix_{-} = 
i(Sx^+ - Cx^-) \rightarrow ix^+$ as $\delta\rightarrow 
\frac{\pi}{4}$.  Therefore the instant defined by $x^+=0$ becomes invariant
under longitudinal boosts as we move to the light-front.
Besides this new feature, the operators in each group change continuously 
as we move from the
equal-time quantization to the light-front.  

The second feature to note is the smooth exchange of the roles of 
transverse boosts and rotations.  In the equal time case ($\delta=0$),
rotations are kinematic and boosts are dynamic.  On the light
front, however, transverse rotations are dynamic and transverse boosts
are kinematic.  In the interpolating case, the kinematic generators
${\cal K}_1$ and ${\cal K}_2$ are mixtures of boosts ($E_1,E_2$) and
rotations ($F_1,F_2$).  The dynamic generators ${\cal D}_1$ and
${\cal D}_2$ are also mixtures of boosts and rotations.  
The mixing coefficients are smooth functions of
interpolating angle, as displayed in Figure 1.

\begin{figure}
\centerline{
\psfig{figure=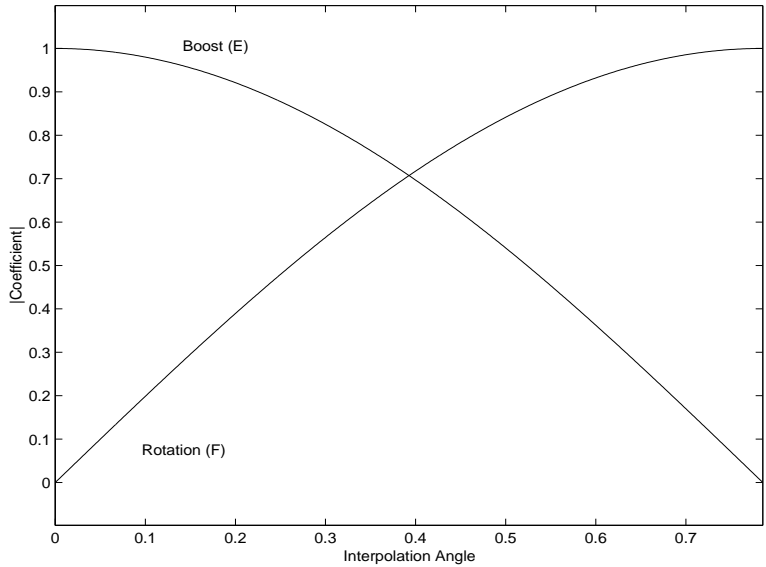,width=4.0in,height=3.0in}
}
\vspace{.1in}
\caption{The smooth exchange of the roles of rotation and boost is
displayed.  The coefficients of rotation $F_i$ and boost $E_i$ in the
dynamic generator ${\cal D}_i$ are plotted versus interpolation
angle $\delta$.}
\label{Dynamic Generator}
\end{figure}
 
We now construct the form for an arbitrary kinematic transformation on
a fixed interpolation front.  In general we have
\begin{equation} \label{T}
T = 
e^{-i\beta_{3}{{K}^{3}}}e^{-i(\beta_{1}{{\cal K}_{1}}+\beta_{2}{{\cal K}_{2}})} 
\end{equation}
where $\beta_1$, $\beta_2$, and $\beta_3$ are free parameters.
Under what conditions is $T$ kinematic?
Use of the Baker-Hausdorff theorem reveals that
\begin{equation}
T^{\dagger}P^{+}T = P^{+}\cosh{\beta_3} + P_{-}\sinh{\beta_3}.
\end{equation}
It follows that $x^+ \rightarrow x^+\cosh{\beta_3} + x_{-}\sinh{\beta_3}$ 
under $T$.  Note that $x^+\cosh{\beta_3} + x_{-}\sinh{\beta_3} = 
x^+\cosh{\beta_3} + ({x^+}S -{x^{-}}C)\sinh{\beta_3}$.
Now $T$ is kinematic if and only if the instant
$x^+=0$ is invariant under $T$.  This requires that 
$-{x^{-}}C\sinh{\beta_3} = 0$.  This can occur only if $C = 0$ or $\beta_3 = 0$.
Thus we find that $T$ is kinematic if $\delta = \frac{\pi}{4}$ or $\beta_3 = 0$.
For $\delta\ne\frac{\pi}{4}$, then, $\beta_3 = 0$ and the kinematic
transformation $T$ has two free parameters.  On the light front,
$\beta_3$ may take on any value and $T$ has three free parameters. 

\section{ SU(2) Spin Algebra and Dynamic Equations in an Arbitrary 
Interpolation Angle}

In this section we construct the SU(2) spin algebra in an arbitrary
interpolation angle.  That is, we wish to construct operators
${\cal J}_i$ satisfying the criteria
\begin{eqnarray}
[{\cal J}_i,{\cal J}_j] &=& i\epsilon_{ijk}{\cal J}_k, \\ \nonumber 
[{\cal J}_i,M] &=& 0, 
\end{eqnarray}
where $M$ is the mass operator.  We also require that $\vec{\cal J}$
commutes with every kinematic generator except $J_3$.  We will see that 
such operators
cannot be defined, in general, on the entire Hilbert space.  Instead,
we define a relevant subspace on which these operators are well-
defined.

\subsection{Kinematic Subspace}

Consider the set of momentum states that can be reached from
rest by a kinematic transformation.  We define this set
of states to be the kinematic subspace for a fixed
interpolating angle.  

In general, kinematic transformations take the form given in Eq. (\ref{T}),
where $\beta_3=0$.
We find that momentum operators transform under $T$ as follows:
\begin{eqnarray}
T^{\dagger}{P_{-}}T &=& P_{-}\cos{\alpha}+\frac{\sin{\alpha}}{\alpha}
C({\beta_1}P^1+{\beta_2}P^2) \nonumber \\
T^{\dagger}{P^{1}}T &=&
P^1-\beta_{1}\frac{\sin{\alpha}}{\alpha}P_{-}+
\frac{\cos{\alpha}-1}{{\alpha}^2}C{\beta_1}({\beta_1}P^1+{\beta_2}P^2) 
\nonumber \\ 
T^{\dagger}{P^{2}}T &=&
P^2-\beta_{2}\frac{\sin{\alpha}}{\alpha}P_{-}+
\frac{\cos{\alpha}-1}{{\alpha}^2}C{\beta_2}({\beta_1}P^1+{\beta_2}P^2), 
\end{eqnarray}
where $\alpha=\sqrt{C({\beta_1}^2+{\beta_2}^2)}$.
This determines how momentum eigenvalues transform under $T$.
At any interpolating angle, the rest state has momentum eigenvalues 
$P^1=P^2=0$ and $P_{-}=-MB$.  It follows that any state which can
be reached from rest must have a three-momentum of the form
\begin{equation} \label{momentum}
(P_{-},P^1,P^2) =
(-MB\cos{\alpha}, \beta_{1}MB\frac{\sin{\alpha}}{\alpha}, 
\beta_{2}MB\frac{\sin{\alpha}}{\alpha}). 
\end{equation}

Suppose we have a momentum state of the form given above.  Then
\begin{equation}
{\bf {P_\perp}}^2 = {P_1}^2+{P_2}^2 = (MB\frac{\sin{\alpha}}{\alpha})^2
({\beta_1}^2+{\beta_2}^2)
\end{equation}
and therefore
\begin{equation} \label{subspace}
{P_{-}}^2 = {M^2}{B^2}-C{\bf {P_\perp}}^2.
\end{equation}
Conversely, any state satisfying (\ref{subspace}) can be reached by a
kinematic transformation.  This condition also implies that $P^+=AM$.

In the equal-time case,
$C=\cos{2\delta}=1$, $B=0$, and $P_{-}=P^3$.  Thus
$(P^3)^2=-{\bf {\bf {P_\perp}}}^2$ and ${\bf P}^2=0$.  The kinematic
subspace in the equal-time case contains only the rest state.  For
a general interpolating angle, the kinematic subspace is a paraboloid
in momentum space containing the origin.
As we move to the light-front limit, $C=0$, $B=-\frac{1}{\sqrt{2}}$,
and $P_{-}=P^{+}$.  Therefore $P^{+}=\frac{M}{\sqrt{2}}$.  In equal-
time momentum space, this is the set of states on the paraboloid
$P^3=-\frac{{\bf {P_\perp}}^2}{2M}$.  If we allow $\beta_3\ne{0}$, however,
the addition of longitudinal boost to the transformation $T$ allows us
to move vertically off of this paraboloid, to a state with arbitrary
longitudinal momentum.  It follows that 
the kinematic subspace on the light-front is identical to the entire
momentum space.  This is a unique feature of $\delta=\frac{\pi}{4}$.

For $0<\delta<\frac{\pi}{4}$ and $P_{\perp}\ne{0}$, $|P_{-}|<-MB$ and 
we can invert Eq. (\ref{momentum}) to find the parameters $\beta_1$ and $\beta_2$.
\begin{eqnarray} \label{betas}
\alpha &=& \arccos{(-\frac{P_{-}}{MB})} \nonumber \\
\beta_1 &=& -\arccos{(-\frac{P_{-}}{MB})}\frac{P^1} 
{\sqrt{M^2{B^2}-{P_{-}}^2}} \nonumber \\
&=& -\arcsin{(\frac{\sqrt{C}|{\bf {P_\perp}}|}{-MB})}\frac{P^1}{|{\bf 
{P_\perp}}|\sqrt{C}} \nonumber \\
\beta_2 &=& -\arccos{(-\frac{P_{-}}{MB})}\frac{P^2}{\sqrt{M^2{B^2}-{P_{-}}^2}}
\nonumber \\
&=& -\arcsin{(\frac{\sqrt{C}|{\bf {P_\perp}}|}{-MB})}\frac{P^2}
{|{\bf {P_\perp}}|\sqrt{C}}.
\end{eqnarray}

\subsection{Construction of SU(2) Algebra}

Following the procedure of Leutwyler and Stern \cite{6}, we now define the spin
operators ${\cal J}_i$ through the use of a kinematic transformation.	
Within the kinematic subspace we define $T$ such that $T|n> = |p,n>$
\footnote{In the equal-time case, the kinematic subspace contains
only the rest state.  Thus we have $T|n>=|n>$.  Since rotations
are kinematic in this case and form an invariant subgroup, $T$ is
not well-defined and may be an arbitrary rotation.  Our goal,
however, is to define helicity in terms of $J^3$ eigenstates $|h>$.  Thus
we require $T|h,n>=|h,n>$, and this forces $T$ to be the identity.  Note
that this ambiguity occurs only in the equal-time case.}.
We define ${\cal J}$ within the subspace such that
\begin{equation}
{\cal J}_i|p,n> = T{J_i}|n>.
\end{equation}
That is, ${\cal J}_i=T{J_i}T^{\dagger}$ on all momentum eigenstates
within the subspace.  Then the operators satisfy the necessary SU(2)
algebra:
\begin{eqnarray}
[{\cal J}_i,{\cal J}_j]|p,n> &=& T{J_i}{J_j}T^{\dagger}|p,n> -
T{J_j}{J_i}T^{\dagger}|p,n> \nonumber \\ 
&=& T[J_i,J_j]T^{\dagger}|p,n> \nonumber \\ 
&=& 
i\epsilon_{ijk}T{J_k}T^{\dagger}|p,n> \nonumber \\ 
&=& i\epsilon_{ijk}{{\cal J}_k}|p,n>.
\end{eqnarray}

Note that the mass operator is defined by
$M^2={P^+}{P_{+}}+{P^-}{P_{-}}-{\bf {P_\perp}}^2$.  Some manipulation
reveals that $[{\cal K}_i,M^2] = 0$, so $[T,M^2] = 0$ and:
\begin{eqnarray}
[{\cal J}_i,M]|p,n> &=& T{J_i}T^{\dagger}M|p,n> - MT{J_i}T^{\dagger}|p,n>
\nonumber \\ 
&=& T{J_i}MT^{\dagger}|p,n> - TM{J_i}T^{\dagger}|p,n> 
\nonumber \\ 
&=&
T({J_i}M-M{J_i})T^{\dagger}|p,n> 
\nonumber \\ 
&=& T({J_i}M-M{J_i})|n> 
\nonumber \\ 
&=& T({J_i}{P^0}-{P^0}{J_i})|n>
\nonumber \\ 
&=& T[J_i,P^0]|n> 
\nonumber \\ 
&=& 0.
\end{eqnarray}

These operators then allow us to define simultaneous eigenstates of
mass and spin.

Finally, each spin operator commutes with $T$ and $T^{\dagger}$:
\begin{eqnarray}
[{\cal J}_i,T^{\dagger}] &=& {\cal J}_iT^{\dagger}|p,n> - 
T^{\dagger}{\cal J}_i|p,n> \nonumber \\ 
&=& {\cal J}_i|n> - T^{\dagger}TJ_i|n> \nonumber \\ 
&=& J_i|n> - J_i|n> = 0.
\end{eqnarray}
It follows that ${\cal J}_i$ commutes with the generators ${\cal K}_1$ and
${\cal K}_2$.  Also, since the action of the spin operators is defined on momentum
eigenstates, it is clear that simultaneous eigenstates of spin and momentum exist.
Thus ${\cal J}_i$ commutes with all three components of the momentum operator.
Thus, the spin operators commute with all generators of the 
stability group except $J_3$, as required.

For a given interpolation angle, we find the angular momentum
operators in terms of the parameters $\beta_1$, $\beta_2$ to be:
\begin{eqnarray} \label{wiggly}
{\cal J}_3 &=& {J_3}\cos{\alpha} + ({\beta_2}{\cal K}_1 - {\beta_1}{\cal K}_2)
\frac{\sin{\alpha}}{\alpha} \nonumber \\ 
{\cal J}_1 &=& J_1 + ({\beta_1}A{J_3} - {\beta_2}B{K_3})
\frac{\sin{\alpha}}{\alpha} \nonumber \\ 
&+& [{{\beta_1}^2}A{\cal K}_2-{{\beta_2}^2}BE_2
-{\beta_1}{\beta_2}(A{\cal K}_1+BE_1)]\frac{\cos{\alpha-1}}{\alpha^2} \nonumber
\\ 
{\cal J}_2 &=& J_2 + ({\beta_2}AJ_3+{\beta_1}BK_3)\frac{\sin{\alpha}}{\alpha} 
\nonumber \\ 
&+& [-{\beta_2}^2A{\cal K}_1 + {\beta_1}^2BE_1 + {\beta_1}{\beta_2}
(A{\cal K}_2 + BE_2)]\frac{\cos{\alpha}-1}{{\alpha}^2},
\end{eqnarray}
where
\begin{eqnarray}
\alpha &=& \sqrt{C({\beta_1}^2+{\beta_2}^2)} \nonumber \\ 
A &=& C\cos{\delta}+S\sin{\delta} \nonumber \\ 
B &=& C\sin{\delta}-S\cos{\delta}. 
\end{eqnarray}

The action of each spin operator is determined by the three
parameters $\delta, \beta_1, \beta_2$.  We first investigate the action
of ${\cal J}$ by fixing $(\beta_1,\beta_2)$ and varying $\delta$.  This
traces a path in momentum space that begins at the rest state and
ends on the surface $P^3=-\frac{{\bf {P_\perp}}^2}{2M}$, as in Fig 2.  
Here three distinct paths are visible, corresponding to different values
for $(\beta_1,\beta_2)$.  Along each path,
the form of ${\cal J}$ is fixed and the coefficient of each generator
depends continuously on interpolation angle.  A path thus connects
states from each kinematic subspace for which the equations of motion
have a fixed form, and allows us to trace the fate of these equations
as we move between interpolation angles.  

In Fig 3, five paths are plotted
in the plane $(P_{\perp},P^3)$ with increasing $\beta_1=\beta_2$.  Connecting
the points on each path parameterized by the same value of $\delta$, we
find a parabola that represents the kinematic subspace for the interpolation
angle $\delta$.  As $\delta$ increases, the parabola opens and becomes
wider.  On the light-front, the endpoints of these paths trace out the
parabola $P^3=-\frac{{\bf {P_\perp}}^2}{2M}$.

\begin{figure}
\centerline{
\psfig{figure=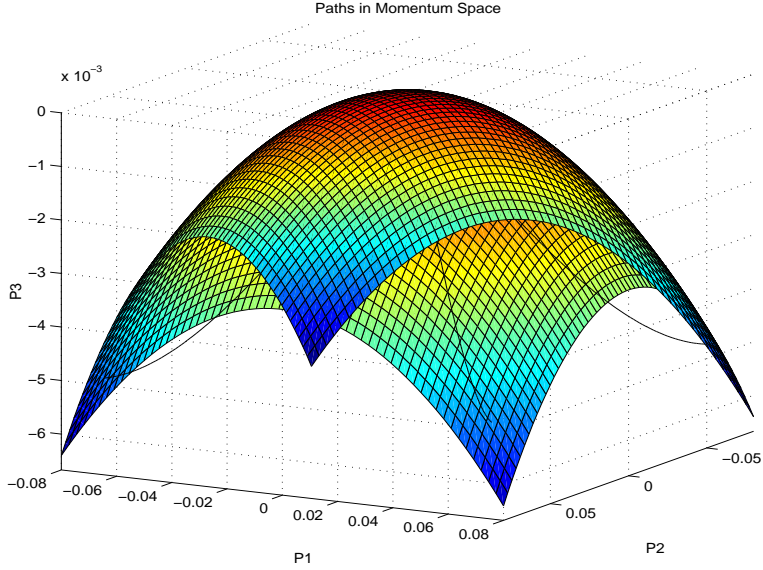,width=4.0in,height=3.0in}
}
\vspace{.1in}
\caption{Three-dimensional plot of momentum states for fixed
$\beta_1,\beta_2$.  Paths parameterized by $\delta$ begin
at the rest state and end on a paraboloid at the light-front.}
\label{Kinematic Paths}
\end{figure}
\begin{figure}
\centerline{
\psfig{figure=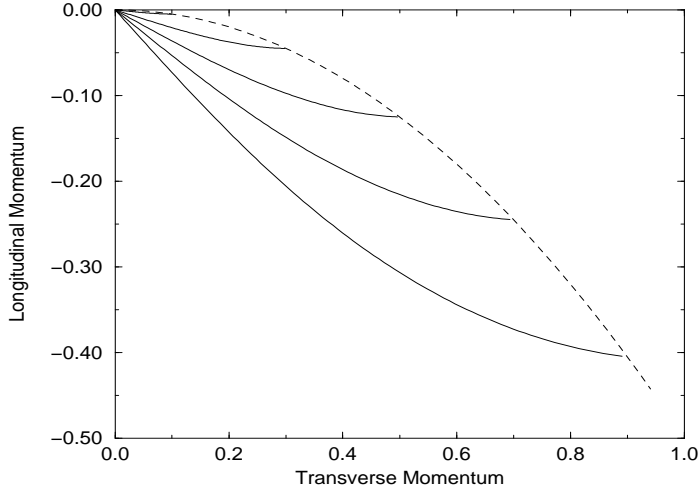,width=4.0in,height=3.0in}
}
\vspace{.1in}
\caption{Momentum states for a fixed $\beta_1=\beta_2$ are plotted
in momentum space.  Each path begins at the rest state and cuts
through a single point on the kinematic subspace.}
\label{Kinematic Paths}
\end{figure}

We now investigate the action of ${\cal J}$ by fixing $\delta$.
For a fixed interpolation angle, we have defined a kinematic
subspace that is parameterized by $\beta_1, \beta_2$.  The action
of the spin operators is defined everywhere on the subspace in
terms of these two parameters.  The representation of ${\cal J}$ at a
fixed $\delta$ 
requires that we define $\beta_1, \beta_2$ in terms of momentum
operators.  Using Eq. (\ref{betas}) we find
\begin{eqnarray}
{\cal J}_3 &=& \frac{-1}{MB}(J_3P_{-}-{\cal K}_1P^2+{\cal K}_2P^1) 
\nonumber \\
\vec{\cal J}_\perp &=& {\bf {J_{\perp}}} + \frac{1}{MB}[{{\bf P}_\perp}A(J_3 -
\frac{\hat{\bf z}\cdot({{\bf P}_\perp}\times{{\vec{\cal K}}_\perp)}}
{MB-P_-}) + (\hat{\bf z}\times{{\bf P}_\perp})B(K_3 - \frac{{{\bf P}_\perp}\cdot
{{\bf E}_\perp}}{MB-P_-})]
\end{eqnarray}
where we may write ${\bf J}_\perp = -\hat{\bf z}\times(A{\vec{\cal K}_\perp}
+B{\vec{\cal D}_\perp})$.
It is straightforward to show that the operator ${\cal J}_3$ commutes with every 
member of the stability group.  We define ${\cal J}_3$ to be the helicity operator.  
Its simple form allows us to trace the fate of helicity states from equal-time to 
the light-front.  The helicity operator can be written in terms of the Pauli-Lubanski 
operator
$W_{\mu} = \frac{1}{2}{\epsilon}_{\mu\nu\alpha\beta}P^{\nu}M^{\alpha\beta}$
as ${\cal J}_3 = \frac{-W^+}{MB}$.

It is important to note that helicity on any quantization front is in general
frame-dependent.  The above property, however, guarantees that helicities are
identical in any two frames that are kinematically connected.  On the light-front,
for example, the Drell-Yan-West and Breit $(q^+=0)$ frames are kinematically
connected.  Thus, helicities must be identical in these two frames as demonstrated
in Ref.\cite{chad}.

For convenience, we present in Appendix B spin-1 and spin-1/2 representations 
for the helicity operator on an arbitrary interpolation front, as well as
spin-1 polarization vectors and Dirac spinors.

\subsection{Limiting Cases}

In Eq. (\ref{betas}), it is clear that problems arise when $\delta=0$, 
$\delta=\frac{\pi}{4}$, or $P_{\perp}=0$.  We now investigate these
problem points and discuss the equal-time and light-front limits.  
First, consider $\beta_1,\beta_2$ as functions of momentum.  
Define $x=\frac{\sqrt{C}|\bf {\bf {P_\perp}}|}{-MB}$, so $0\le{x}\le{1}$.  We
expand $\beta_1,\beta_2$ in powers of $x$.  For $x<<1$ we keep the first term
to find
\begin{eqnarray}
\beta_1=-\frac{xP^1}{|{\bf {P_\perp}}|\sqrt{C}}=\frac{P^1}{MB} \nonumber \\
\beta_2=-\frac{xP^2}{|{\bf {P_\perp}}|\sqrt{C}}=\frac{P^2}{MB}. 
\end{eqnarray}
For $\delta\ne{0}$, we find $\beta_i\rightarrow{0}$ as 
$|P_{\perp}|\rightarrow{0}$.  On the rest state, then, the action of
${\cal J}_i$ is identical to the action of the equal-time angular
momentum $J_i$.  It follows that $\beta_1,\beta_2$ can be
defined as continuous functions of momentum everywhere 
on the kinematic subspace.

Now let us consider the spin operators on the light-front.  Recall that our construction
of a general kinematic transformation $T$ required that $\beta_3=0$, which is necessary
for $0\le\delta<\frac{\pi}{4}$.  In the special case of the light-front, however,
the appearance of $K_3$ as a kinematic operator allows us to define a general kinematic 
transformation with $\beta_3\ne{0}$.  Since the entire momentum space may be 
parameterized by $\beta_1,\beta_2,\beta_3$, 
the kinematic subspace becomes the entire momentum space.  
This is a unique and important feature of the light-front quantization 
$\delta=\frac{\pi}{4}$.

It follows that the action of the spin operators can be defined on any momentum eigenstate.
The operators may be obtained as before, now using the kinematic transformation $T$ with
$\beta_3=\ln\frac{P^+\sqrt{2}}{M}$, $\beta_{\perp{i}}=-\frac{P^i}{P^+}$ 
\cite{6}.  The resulting spin operators, 
valid for any momentum state, are given in terms of $\beta_1,\beta_2,\beta_3$:
\begin{eqnarray}
{\cal J}_3 &=& J_3+{\beta_1}E_2-{\beta_2}E_1 \nonumber \\ 
{\cal J}_1 &=& \frac{1}{\sqrt{2}}[({\beta_1}J_3+{\beta_2}K_3+
\frac{1}{2}({\beta_1}^2-{\beta_2}^2)E_2
+{\beta_1}{\beta_2}E_1)e^{\beta_3} - 
{E_2}e^{-\beta_3}+{F_2}e^{\beta_3}] \nonumber \\ 
{\cal J}_2 &=& \frac{1}{\sqrt{2}}[({\beta_2}J_3-
{\beta_1}K_3+\frac{1}{2}({\beta_2}^2-{\beta_1}^2)E_2
-{\beta_1}{\beta_2}E_1)e^{\beta_3} - 
{E_2}e^{-\beta_3}+{F_2}e^{\beta_3}]. 
\end{eqnarray}

The light-front spin operators are therefore
\begin{eqnarray}\label{LF-angula}
{\cal J}_3 &=& J_3 + \frac{1}{P^+}({P^2}E_1-{P^1}E_2) \nonumber \\
{\cal J}_1 &=& \frac{1}{M}({P^+}F_2-{P^-}E_2-{P^2}K_3-{P^1}{\cal J}_3) 
\nonumber \\ 
{\cal J}_2 &=& \frac{1}{M}({P^-}E_1-{P^+}F_1+{P^1}K_3-{P^2}{\cal J}_3). 
\end{eqnarray}
These are the spin operators presented in Appendix B of 
Ref.\cite{keister},
where the operators $P^+,E_i,F_i$ do not contain our normalization factor of
$\frac{1}{\sqrt{2}}$. 
It is important to compare this general light-front result with our 
interpolating spin operators.

Consider the light-front limit of the spin operators given in Eq. 
(\ref{wiggly}).
In the light-front limit $C\rightarrow{0}$, and using the previous expansion
we find that $\beta_1\rightarrow
\frac{P^1}{MB}=-\frac{P^1\sqrt{2}}{M}$.  The spin operators become
\begin{eqnarray}\label{LF-J}
{\cal J}_3 &=& J_3 + \frac{\sqrt{2}}{M}(P^2{E_1}-P^1{E_2}) \nonumber \\ 
{\cal J}_1 &=& 
\frac{1}{\sqrt{2}}({F_2}-{E_2})-\frac{1}{M}(P^1{J_3}+P^2{K_3}) \nonumber \\
&-& \frac{1}{{M^2}\sqrt{2}}[(-(P^1)^2+(P^2)^2){E_2}+2P^1{P^2}E_1] 
\nonumber \\ 
{\cal J}_2 &=& \frac{1}{\sqrt{2}}({E_1}-{F_1})-\frac{1}{M}(P^2{J_3}-P^1{K_3}) 
\nonumber \\
&-& \frac{1}{{M^2}\sqrt{2}}[(-(P^1)^2+(P^2)^2){E_1}-2P^1{P^2}E_2].
\end{eqnarray}
Recall that in the light-front limit, ${\cal J}$ is defined on the subspace
$\sqrt{2}{P^+}=M$.  Within this subspace, Eq.(\ref{LF-J}) coincides with 
Eq.(\ref{LF-angula}).
It follows that our interpolating spin operators 
are consistent
with the light-front spin operators within the subspace on which they 
are defined.

Now let us consider the equal-time limit.  
In the equal-time limit the kinematic subspace contains only the rest state, and
$T$ becomes a rotation.  Since any
sequence of rotations leaves the rest state invariant, the parameters $\beta_1,\beta_2$ 
may take on any value (see footnote 3).  The spin operators become
\begin{eqnarray}
{\cal J}_3 &=& J_3\cos{\alpha}+(-\beta_2{J_2}-\beta_1{J_1})
\frac{\sin{\alpha}}{\alpha} \nonumber
\\ 
{\cal J}_1 &=& J_1 + {\beta_1}{J_3}\frac{\sin{\alpha}}{\alpha}
+ ({\beta_1}^2{J_1}+{\beta_1}{\beta_2}J_2)\frac{\cos{\alpha}-1}{\alpha^2} 
\nonumber \\ 
{\cal J}_2 &=& J_2 + {\beta_2}{J_3}\frac{\sin{\alpha}}{\alpha}
+ ({\beta_2}^2{J_2}+{\beta_1}{\beta_2}J_1)\frac{\cos{\alpha}-1}{\alpha^2},
\end{eqnarray}
where $\alpha = \sqrt{{\beta_1}^2+{\beta_2}^2}$.  If we set $\beta_1=0$, then
\begin{eqnarray}
{\cal J}_3 &=& {J_3}\cos{\beta_2}-{J_2}\sin{\beta_2} \nonumber \\ 
{\cal J}_1 &=& J_1 \nonumber \\ 
{\cal J}_2 &=& {J_2}\cos{\beta_2}+{J_3}\sin{\beta_2}. 
\end{eqnarray}
These are rotated about the x-axis.
Similarly if we set $\beta_2=0$, then  
\begin{eqnarray}
{\cal J}_3 &=& {J_3}\cos{\beta_1}-{J_1}\sin{\beta_1} \nonumber \\ 
{\cal J}_1 &=& {J_1}\cos{\beta_1}+{J_3}\sin{\beta_1} \nonumber \\ 
{\cal J}_2 &=& J_2.
\end{eqnarray}
These are rotated about the y-axis.  
Thus the spin operators are angular momentum operators $J_1,J_2,J_3$, 
about rotated axes, as we should
expect.  When we restrict $T$ to be the identity transformation,
each $\beta_i=0$, and we recover the ordinary equal-time
angular momentum operators.

\section{ Discussion and Conclusion }

In this work, we constructed the Poincare algebra valid for any 
interpolation angle between the instant limit and the light-front limit.
We find that the light-front limit $\delta=\frac{\pi}{4}$ is a special  
angle that adds a new kinematic operator $K_3$. The conversion of
the dynamical operator between boost and rotation is quite smooth as
shown in Fig.1. The general result of ${\cal J}$ in the instant limit
agrees with the ordinary angular momentum ${\vec J}$ while it agrees
with the LF ${\cal J}$ obtained by Leutwyler and Stern in the light-front
limit with operation to the parabolic subspace shown in Fig.2. 
It is interesting to note that the subspace is limited to only the
rest frame in the instant limit while it can expand to an arbitrary frame 
in the light-front limit. We have also presented the helicity operator
in an arbitrary interpolation angle. Explicit verification for the 
correct helicity eigenvalues is summarized in the Appendix B.
Since our results are model independent, they can play the role of
testing any suggested hadron model. Our results indicate that the  
interpolation method preserving the orthogonal coordinate system
is useful in tracing the fate of interesting results obtained by
one form of Hamiltonian dynamics in the other end of interpolation angle.
Applications to other nonperturbative analyses such as the BCS vacuum
and the mass gap equation are under consideration.

\acknowledgements
\noindent
This work was supported in part by a grant from the US Department of
Energy under contracts DE-FG02-96ER40947. The North Carolina
Supercomputing Center and the National Energy Research Scientific
Computer Center are also acknowledged for the grant of supercomputer time.

\newpage
\noindent
\appendix
\setcounter{section}{0}
\setcounter{equation}{0}
\setcounter{figure}{0}
\renewcommand{\theequation}{\mbox{A\arabic{equation}}}
\section{ Poincar{\'e} Algebra on an Arbitrary Interpolating Front}

\begin{eqnarray}
[P^-,P^1] &=& 0, \\ \nonumber 
[P^-,P^2] &=& 0, \\ \nonumber
[P^-,P^+] &=& 0, \\ \nonumber
[P^-,E_2] &=& iP^2{S}, \\ \nonumber
[P^-,E_1] &=& iP^1{S}, \\ \nonumber
[P^-,J^3] &=& 0, \\ \nonumber	
[P^-,F_1] &=& -iP^1{C}, \\ \nonumber
[P^-,F_2] &=& -iP^2{C}, \\ \nonumber
[P^-,K^3] &=& -iP_{+}, \\ \nonumber
[P^1,P^2] &=& 0, \\ \nonumber 
[P^1,P^+] &=& 0, \\ \nonumber
[P^1,E_2] &=& 0, \\ \nonumber
[P^1,E_1] &=& iP^+ \\ \nonumber
[P^1,J^3] &=& -iP^2 \\ \nonumber
[P^1,F_1] &=& iP^{-} \\ \nonumber
[P^1,F_2] &=& 0 \\ \nonumber
[P^1,K^3] &=& 0 \\ \nonumber
[P^2,P^+] &=& 0 \\ \nonumber
[P^2,E_2] &=& iP^+ \\ \nonumber
[P^2,E_1] &=& 0 \\ \nonumber
[P^2,J^3] &=& iP^1 \\ \nonumber
[P^2,F_1] &=& 0 \\ \nonumber
[P^2,F_2] &=& iP^- \\ \nonumber
[P^2,K^3] &=& 0 \\ \nonumber
[P^+,E_2] &=& iP^2{C} \\ \nonumber
[P^+,E_1] &=& iP^1{C} \\ \nonumber
[P^+,J^3] &=& 0 \\ \nonumber
[P^+,F_1] &=& iP^1{S} \\ \nonumber
[P^+,F_2] &=& iP^2{S} \\ \nonumber
[P^+,K^3] &=& iP_{-} \\ \nonumber
[E_2,E_1] &=& iJ^3{C} \\ \nonumber
[E_2,J^3] &=& iE_1 \\ \nonumber
[E_2,F_1] &=& iJ^3{S} \\ \nonumber
[E_2,F_2] &=& -iK^3 \\ \nonumber
[E_2,K^3] &=& -i{\cal K}_2 \\ \nonumber
[E_1,J^3] &=& -iE_2 \\ \nonumber
[E_1,F_1] &=& -iK^3 \\ \nonumber
[E_1,F_2] &=& -iJ^3{S} \\ \nonumber
[E_1,K^3] &=& -i{\cal K}^1 \\ \nonumber
[J^3,F_1] &=& iF_2 \\ \nonumber
[J^3,F_2] &=& -iF_1 \\ \nonumber
[J^3,K^3] &=& 0 \\ \nonumber
[F_1,F_2] &=& iJ^3{C} \\ \nonumber
[F_1,K^3] &=& i{\cal D}^1 \\ \nonumber
[F_2,K^3] &=& i{\cal D}^2
\end{eqnarray}

\newpage
\noindent
\section{ Helicities on an Arbitrary Interpolation Front}
\renewcommand{\theequation}{\mbox{B\arabic{equation}}}

The helicity operator may be represented for spin-1/2 states
as follows:
\begin{equation}
[{\cal J}_3]_{\frac{1}{2}} = -\frac{1}{MB}
\left(\begin{array}{cccc} P_{-} & AP_L & 0 & BP_L  \\ AP_R & -P_{-} & -BP_R & 0 \\ 
0 & BP_L & P_{-} & AP_L \\ -BP_R & 0 & AP_R & -P_{-} \end{array} \right). \\ \nonumber
\end{equation}
Here $P_R = P^1 + iP^2$ and $P_L = P^1 - iP^2$.
We found the spin-1/2 eigenstates of helicity by diagonalizing this matrix.  
These spinors are the solutions of the Dirac equation for an arbitrary 
interpolation angle: 
\begin{eqnarray}
u(p,+1) &=& -\frac{1}{\sqrt{B(MB-P_-)}}\left(\begin{array}{c} MB-P_-  \\ -AP_R \\ 
0  \\ BP_R \end{array} \right) \\ \nonumber
u(p,-1) &=& -\frac{1}{\sqrt{B(MB-P_-)}}\left(\begin{array}{c} AP_L \\ MB-P_- \\ 
BP_L \\ 0 \end{array} \right). \\ \nonumber
\end{eqnarray}

In addition, these spin-1/2 eigenstates are antiparticle solutions to the Dirac
equation for an arbitrary angle:
\begin{eqnarray}
v(p,+1) &=& \frac{1}{\sqrt{B(MB-P_-)}}\left(\begin{array}{c} BP_L  \\ 0 \\ 
AP_L  \\ MB-P_- \end{array} \right) \\ \nonumber
v(p,-1) &=& -\frac{1}{\sqrt{B(MB-P_-)}}\left(\begin{array}{c} 0 \\ BP_R \\ 
MB-P_- \\ -AP_R \end{array} \right). \\ \nonumber
\end{eqnarray}

These spinors satisfy the following constraints:
\begin{eqnarray}
({\not\! P} - m)u(p,\lambda) &=& 0 \\ \nonumber
({\not\! P} + m)v(p,\lambda) &=& 0 \\ \nonumber
\bar{u}(p,\lambda)u(p,\lambda') &=& 2m{\delta}_{\lambda\lambda'} 
= -\bar{v}(p,\lambda')v(p,\lambda) \\ \nonumber
{\sum_\lambda}u(p,\lambda)\bar{u}(p,\lambda) &=& {\not\! P} + M \\ \nonumber
{\sum_\lambda}v(p,\lambda)\bar{v}(p,\lambda) &=& {\not\! P} - M \\ \nonumber
\end{eqnarray}

Similarly, for spin-1 states we have the representation
\begin{equation}
[{\cal J}_3]_1 = \frac{i}{MB}
\left(\begin{array}{cccc} 0 & -BP^2 & BP^1 & 0  \\ -BP^2 & 0 & -P_{-} & AP^2 \\ 
BP^1 & P_{-} & 0 & -AP^1 \\ 0 & -AP^2 & AP^1 & 0 \end{array} \right). \\ \nonumber
\\ \nonumber
\end{equation}
We found the spin-1 eigenstates of helicity by diagonalizing this matrix.  After
proper normalization, we obtain the polarization vectors given by 
\begin{eqnarray}
\epsilon(p,0) &=& -\frac{A}{MB}(P_{+}-\frac{M^2}{P^+},P_{-},P^1,P^2)
\\ \nonumber
\epsilon(p,+1) &=& \frac{1}{\sqrt{2}MB}(S|P_{\perp}|, -C|P_{\perp}|,
\frac{P_{-}P^1-iMBP^2}{|P_{\perp}|},
\frac{P_{-}P^2+iMBP^1}{|P_{\perp}|})
\\ \nonumber
\epsilon(p,-1) &=& \frac{1}{\sqrt{2}MB}(S|P_{\perp}|, -C|P_{\perp}|,
\frac{P_{-}P^1+iMBP^2}{|P_{\perp}|},
\frac{P_{-}P^2-iMBP^1}{|P_{\perp}|}),
\\ \nonumber
\end{eqnarray}
where $\epsilon(p,\lambda)$ is written in the form
$\epsilon(p,\lambda) = (\epsilon_+,\epsilon_-,\epsilon_1,\epsilon_2,)$.
  These polarization vectors satisfy the constraints
\begin{eqnarray}
\epsilon(p,\lambda)\cdot{p} &=& 0 \\ \nonumber
\epsilon^{*}(p,\lambda)\cdot\epsilon(p,{\lambda}') &=& -\delta_{\lambda\lambda'} \\ \nonumber
\sum_{\lambda}\epsilon^{\mu}(p,\lambda)\epsilon^{\nu}(p,\lambda) &=&
-g^{\mu\nu}+\frac{p^{\nu}p^{\mu}}{M^2}. \\ \nonumber
\end{eqnarray}
It is also clear that the longitudinal polarization vector ${\epsilon}(p,0)$
is "parallel" to the three-momentum $\bf P$, since $(\epsilon_-,\epsilon_1,
\epsilon_2)\sim{(P_-,P_1,P_2)}$.  This is a feature of both light-front and
traditional equal-time definitions of longitudinal helicity $(h=0)$.

\end{document}